\documentclass[conference,twocolumn]{IEEEtran}
\usepackage{setspace}
\usepackage{enumerate}
\usepackage{graphicx}
\usepackage{epstopdf}
\usepackage{csquotes}
\usepackage{subfig}
\usepackage{mathtools}
\usepackage{amssymb, amsmath,amsthm, amsfonts}
\usepackage{ifthen}
\usepackage{tikz}
\usepackage{enumerate}
\usepackage{verbatim}
\usepackage{pifont}
\usepackage{bm}  
\usepackage{stackengine}
\usepackage{bbm}

\DeclareSymbolFont{bbold}{U}{bbold}{m}{n}
\DeclareSymbolFontAlphabet{\mathbbold}{bbold}

\newtheorem{theorem}{Theorem}

\newtheorem{lemma}[theorem]{Lemma}

\newtheorem{remark}[theorem]{Remark}

\newcommand{\be}{\begin{equation}}
\newcommand{\ee}{\end{equation}}
\newcommand{\ben}{\begin{equation*}}
\newcommand{\een}{\end{equation*}}
\newcommand{\ba}{\begin{eqnarray}}
\newcommand{\ea}{\end{eqnarray}}

\newcommand{\ind}{\mathbbm 1}
\newcommand{\Y}{\mathcal{Y}}
\newcommand{\U}{\mathcal{U}}
\newcommand{\W}{\mathcal{W}}
\newcommand{\V}{\mathcal{V}}
\newcommand{\X}{\mathcal{X}}
\newcommand{\M}{\mathcal{M}}

\newcommand{\Z}{\mathcal{Z}}

\newcommand{\Lcal}{\mathcal{L}}

\newcommand\firstequ{\mathrel{\overset{\makebox[0pt]{\mbox{\normalfont\tiny\sffamily (a)}}}{=}}}

\newcommand\secondequ{\mathrel{\overset{\makebox[0pt]{\mbox{\normalfont\tiny\sffamily (b)}}}{=}}}

\newcommand\firstineq{\mathrel{\overset{\makebox[0pt]{\mbox{\normalfont\tiny\sffamily (c)}}}{\le}}}

\newcommand{\suppress}[1]{}

\def\h2{\tilde h}

\def\hm1{\hat h_{-1}}

\begin{document}

\title{ Distributed Hypothesis  Testing Over  Noisy Channels}
\author{Sreejith Sreekumar and Deniz G\"und\"uz \\ Imperial College London, UK \\
Email: \{s.sreekumar15, d.gunduz\}@imperial.ac.uk}  
\maketitle

\begin{abstract}
 A distributed binary hypothesis testing  problem,  in which  multiple observers transmit their observations to a detector over noisy channels, is studied. Given its own side information, the goal of the detector is to decide between  two hypotheses for the joint distribution of the data. 
Single-letter upper and lower  bounds on  the optimal type 2 error exponent (T2-EE), when the type 1 error probability vanishes with the block-length are obtained. These bounds coincide and characterize the optimal T2-EE when only a single helper is involved. Our result shows that the optimal T2-EE depends on the marginal distributions of the data and the channels rather than  their joint distribution. However, an operational separation between  HT and channel coding does not hold, and the optimal T2-EE is achieved by generating channel inputs correlated with  observed data.   
\end{abstract}

\section{Introduction}

 
 Statistical inference and learning have assumed prime importance  in the fields of machine learning, data analytics and communications applications. An important problem arising in these scenarios is that of discerning the statistics of the available data. This leads to the formulation of a hypothesis  testing (HT) problem, in which  the objective  is to identify the underlying probability distribution of the data samples, from among a  set of candidate  distributions. With the increasing adaption of distributed sensing technologies and the Internet of Things (IoT) paradigm, the data is often  collected from multiple remote locations and   communicated to the  detector over noisy communication links. This naturally leads to the   problem of  distributed statistical inference over noisy communication channels. 

\begin{figure}[t]
\centering
\includegraphics[height=3.4cm, width=8cm]{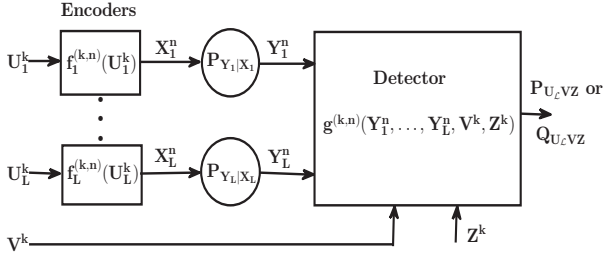}
\caption{Illustration of a distributed hypothesis testing system over noisy channels} \label{htnoisymodel}
\end{figure}

In this paper, we study the problem of  distributed binary HT over  noisy channels depicted in Fig. \ref{htnoisymodel}. The detector is interested in determining whether the data  $(U_1,\ldots,U_L,V,Z)$ is distributed according to $P_{U_1 \ldots U_LVZ}$ or $Q_{U_1 \ldots U_LVZ}$ under hypotheses $H_0$ and $H_1$, respectively. Each encoder $l$, $l=1, \ldots, L$, observes $k$ samples independent and identically distributed (i.i.d) according to $P_{U_l}$, and  communicates its observation to the detector by $n$ uses of the discrete memoryless channel (DMC), characterized by the conditional distribution $P_{Y_l|X_l}$. The detector decides between the two hypotheses $H_0$ and $H_1$ based on the channel outputs $Y_1^n, \ldots,Y_L^n$ as well as its own observations $V^k$ and $Z^k$.    Our goal is to  characterize  the optimal type 2 error exponent (T2-EE) for this model as a function of the  bandwidth ratio, $\tau=\frac{n}{k}$,  under the constraint that the type 1 error probability is less than a specified value. We will  focus mostly on the special case in which $P_{U_1 \ldots U_LVZ}= P_{U_1 \ldots U_LV|Z}P_Z$ and  $Q_{U_1  \ldots U_LVZ}= P_{U_1 \ldots U_L|Z}P_{V|Z}P_Z$, known as the \textit{testing against conditional  independence (TACI)}  problem. 

Distributed statistical inference under communication constraints was originally formulated by Berger in \cite{Berger_1979}. A simplified version of  this is considered in  \cite{Ahlswede-Csiszar}, which   studies binary HT for the model in Fig. \ref{htnoisymodel} when $L=1$, $Z$ is absent and the channel between the encoder and the  detector is a noise-free channel of rate $R$. Ahlswede and Csisz\'{a}r establish a single-letter characterization of the optimal T2-EE for the  testing against independence (TAI) problem (including a strong converse), along with single-letter lower bounds  for  the general HT problem in \cite{Ahlswede-Csiszar}. 
 For the same model, \cite{Han} provides a tighter lower bound on the T2-EE, which coincides with that of \cite{Ahlswede-Csiszar} for the   TAI problem. An improved lower bound for the same problem is obtained in \cite{Shimokawa} by introducing \enquote{binning} at the encoder. HT for the model in    Fig. \ref{htnoisymodel} with  noise free rate-limited channels is studied  in \cite{Rahman-Wagner}, and  the authors  establish the optimality of binning  for the  TACI problem. A single-letter characterization of  the optimal T2-EE  for the multi-terminal  TAI problem  is  obtained in  \cite{Zhao-Lai} under a certain Markovian condition. In a slightly different setting with two decision centers, the optimal T2-EE  for a three terminal dependence testing problem is characterized in  \cite{Wigger-Timo}. The optimal T2-EE, when multiple interactions between the encoder and detector are allowed, is studied in  \cite{Katz},\cite{Xiang-Kim}. We remark here that all the above works consider rate-limited bit-pipes  from the observers to the detector, and to the best of our knowledge, HT over noisy channels has not been studied previously. 
 
\textbf{Notations:} The  support of  a random variable (r.v.) is denoted by calligraphic letters, e.g., $\X$ for r.v. $X$. The cardinality of $\X$ is denoted by $|\X|$. The joint distribution of r.v.'s $X$ and $Y$ is denoted by $P_{XY}$ and its marginals by $P_X$ and $P_Y$.  $X-Y-Z$ denotes that $X,~Y,~Z$ form a Markov  chain. For $m \in \mathbb{Z}^+$, $X^m$ denotes the sequence $X_1, \ldots, X_m$, while $X_{l}^m$ denotes $X_{l,1}, \ldots, X_{l,m}$ associated with observer $l$. The group of $m$ r.v's $X_{l,((j-1)m+1)}, \ldots, X_{l,((jm)}$ is denoted by   $X_l^m(j)$, and   
  the infinite sequence $ X_l^m(1), X_l^m(2), \ldots$  is denoted by  $\{X_l^m(j)\}_{j \in \mathbb{Z}^+}$. Similarly, for a subset $S=\{l_1, \ldots, l_{s}\}$ of observers,   $\left\lbrace X_{l_1}^m, \ldots, X_{l_{s}}^m\right\rbrace$, $\left\lbrace X_{l_1}^m(j), \ldots, X_{l_{s}}^m(j)\right\rbrace$  and $\left\lbrace \left\lbrace X_{l_1}^m(j) \right\rbrace _{ j \in \mathbb{Z}^+},\ldots, \left\lbrace X_{l_{s}}^m(j) \right\rbrace _{ j \in \mathbb{Z}^+} \right\rbrace$  are  denoted by  $X_{S}^m$, $X_{S}^m(j)$ and $\left\lbrace X_S^m(j)\right\rbrace_{j \in  \mathbb{Z}^+}$, respectively. 
   Following the notation in \cite{Csiszar-Korner}, $T_{P}$ and $T_{[X]_{\delta}}^m$ (or  $T_{\delta}^m$ when there is no ambiguity) denote the set of sequences of type $P$ and the set of $P_X-$ typical sequences of length $m$, respectively. 
  $D(P||Q)$ denotes the  Kullback-Leibler (KL) divergence between distributions $P$ and $Q$ \cite{Csiszar-Korner}. 
   All logarithms  are to the base 2. $\ind$ denotes the indicator function.

\section{System Model}
All the r.v.'s considered henceforth are discrete with finite support. Let $k,n \in \mathbb{Z}^+$ be arbitrary.  Let  $\Lcal=\{1, \ldots,L\}$ denote the set of observers which communicate to the detector over orthogonal noisy channels, as shown in Fig. \ref{htnoisymodel}.  For $l \in \Lcal$, encoder $l$ observes   $U_{l}^k$  and transmits  $X_l^n= f_l^{(k,n)}(U_l^k)$, where $f_l^{(k,n)}: \U_l^k \rightarrow \X_l^n$ is a stochastic mapping.
Let  $\tau \triangleq \frac{n}{k}$ denote the \textit{bandwidth ratio}. 
The channel  output $Y_{\Lcal}^n$ is given by  the probability law $P_{Y_{\Lcal}^n| X_{\Lcal}^n }(y_{\Lcal}^n|x_{\Lcal}^n)= \prod_{l=1}^L\prod_{j=1}^n P_{Y_l|X_l}(y_{l,j}|x_{l,j})$, i.e., the channels between the  observers  and  the detector are orthogonal and discrete memoryless.
Depending on the received symbols  $Y_{\Lcal}^n$ and samples $(V^k,Z^k)$, the detector makes a decision  between the two hypotheses $H_0:P_{U_{\Lcal}VZ}$ or $H_1:Q_{ U_{\Lcal} V Z}$ according to the  decision rule $g^{(k,n)}:\Y_{\Lcal}^n \times \V^k \times \Z^k \rightarrow \{0,1\} $ given by $g^{(k,n)}(y_{\Lcal}^n,v^k,z^k)= \ind \left( (y_{\Lcal}^n,v^k,z^k) \in A^c\right)$,
where $A$ denotes the acceptance region for $H_0$.  It is assumed that the r.v's $U_{\Lcal},V$ and $Z$ have the same marginal distributions under both $H_0$ and $H_1$  and that $Q_{U_{\Lcal}VZ}(u_{\Lcal},v,z)>0 $ for all $( u_{\Lcal}, v,z) \in \U_{\Lcal} \times \V \times \Z$. In this paper, we  focus  mostly on  the special case when 
$H_0:P_{U_{\Lcal}V|Z}P_Z$ and  $H_1: P_{ U_{\Lcal}|Z}P_{V|Z}P_Z$, i.e, TACI between $V$ and $U_{\Lcal}$ given $Z$.


Let  $\bar \alpha \left(k,n, f_{1}^{(k,n)}, \ldots,f_{L}^{(k,n)}, g^{(k,n)} \right)\triangleq P_{Y_{\Lcal}^{n} V^{k} Z^{k}}(A^c)$ and $\bar \beta \left(k,n, f_{1}^{(k,n)}, \ldots, f_{1}^{(k,n)}, g^{(k,n)} \right)\triangleq Q_{Y_{\Lcal}^{n} V^{k} Z^{k}}(A)$ denote the type 1 and type 2  error probabilities, respectively. Define
\begin{align}
&\beta' \left(k,n, f_{1}^{(k,n)}, \ldots,f_{L}^{(k,n)}, \epsilon \right)  \triangleq \nonumber\\  
&\qquad \quad \quad \inf_{g^{(k,n)}} \bar \beta \left(k,n, f_{1}^{(k,n)}, \ldots,f_{L}^{(k,n)},~g^{(k,n)} \right) \label{deft2error}
\end{align}
such that 
\begin{subequations}
\begin{equation}
\bar \alpha \left(k,n, f_{1}^{(k,n)}, \ldots, f_{L}^{(k,n)},~g^{(k,n)} \right) \leq  \epsilon,
\end{equation}
\begin{equation}
(Z^k,~V^k) - U_l^k-X_l^n = f_l^{(k,n)}(U_l^k)- Y_l^n, ~l \in \Lcal,
\end{equation}
\end{subequations}
and let 

\begin{equation}
\beta(k, \tau,\epsilon) \triangleq \inf_{\substack{f_{1}^{(k,n)}, \ldots,f_{L}^{(k,n)}, \\ n \leq \tau k}} \beta' \left(k,n, f_{1}^{(k,n)}, \ldots,f_{L}^{(k,n)},\epsilon \right). 
\end{equation}
%
%
Note that $\beta(k, \tau,\epsilon)$ is a non-increasing  function of $k$ and $\epsilon$.
A T2-EE $\kappa'$ is said  to be   $(\tau, \epsilon)$ achievable  if there exists  a sequence of  integers $k$, encoding functions $f_l^{(k,n_k)}: \U^{k} \rightarrow \X^{n_k}, ~ l \in \Lcal$ and  decoding function $g^{(k,n)}$  such that $n_k \leq \tau k$, $\forall~k$, and  for any $\delta>0$,

\begin{align} 
\limsup_{k \rightarrow \infty} \frac{\log \left( \beta(k, \tau,\epsilon) \right)}{k} & \leq -(\kappa'- \delta).
 \label{seqkappa}
\end{align}
Let $\kappa(\tau,\epsilon )  \triangleq  \sup \{\kappa': \kappa' \text{ is  } (\tau,\epsilon) \text{ achievable}\}$.

For $k \in \mathbb{Z}^+$, we define
\begin{align}
\theta (k,\tau) &\triangleq  \sup_{ \substack{f_1^{(k,n)},\ldots,f_L^{(k,n)} \\ n \leq \tau k}} 
     \frac{D \big(P_{ Y_{\Lcal}^n V^kZ^k }|| Q_{ Y_{\Lcal}^n  V^k  Z^k } \big)}{k}, \label{thetaktau} 
\end{align}
and 
\begin{align}
 \theta (\tau)& \triangleq \sup_{k } \theta(k, \tau). \label{thetataudef}
\end{align}

%
In this paper, we obtain  single-letter  upper  and lower  bounds on  $\kappa( \tau,\epsilon)$ for the TACI problem. It is shown that the two bounds coincide  when $L=1$. Our approach is similar to that in \cite{Ahlswede-Csiszar}, where we first obtain bounds for $\kappa( \tau,\epsilon)$ in terms of $\theta$, and then show that $\theta$ has a single-letter characterization in terms of information theoretic quantities. We establish this characterization  by considering the  joint  source-channel coding (JSCC) problem with noisy helpers.  
The next lemma obtains the bounds for $\kappa( \tau,\epsilon)$ in terms of $\theta$.

\begin{lemma} \label{lem:achandconverse}
For any  bandwidth ratio $\tau>0$, we have 
\begin{enumerate}[(i)]
\item  $\limsup_{k \rightarrow \infty  }\frac{ \log \left(\beta(k, \tau,\epsilon)\right)}{k}  \leq - \theta(\tau)$, 
$\epsilon \in (0,1)$.
\item $ \lim_{\epsilon \rightarrow 0}\liminf_{ k \rightarrow \infty } \log \left(\frac{\beta(k,\tau, \epsilon)}{k} \right) \geq - \theta(\tau)$.
\end{enumerate}
\end{lemma}

\begin{IEEEproof}
The proof is similar to  that  of  Theorem 1 in \cite{Ahlswede-Csiszar}. 
We  prove $(i)$ and  omit the proof of $(ii)$ due to space limitations. 
Let $k \in \mathbb{Z}^+$ and $\tilde \epsilon>0$ be arbitrary, and  $\tilde n_k$, $\tilde f_l^{(k,\tilde n_k)}, ~ l \in \Lcal,$ and $ \tilde Y_{\Lcal}^{\tilde n_k} $ be the  channel block length, encoding functions and  channel outputs respectively, such that $k \theta(k,\tau)- D \big(P_{ Y_{\Lcal}^{\tilde n_k} V^kZ^k }|| Q_{ Y_{\Lcal}^{\tilde n_k}  V^k  Z^k } \big) < k \tilde \epsilon$   . 
 For  each $l \in \Lcal$, 
$\left\lbrace\tilde Y_{l}^{\tilde n_k}(j)\right\rbrace_{j \in \mathbb{Z}^+}$  form an infinite sequence of i.i.d. r.v.'s indexed by $j$. 
Hence, by the application of Stein's Lemma \cite{Ahlswede-Csiszar} to the  sequences $\left\lbrace\tilde Y_{\Lcal}^{\tilde n_k}(j), V^{k}(j),Z^{k}(j)\right\rbrace_{j \in \mathbb{Z^+}}$, we have 
\begin{align}
\limsup_{j \rightarrow \infty}  \frac{\log \left(\beta(kj,\tau,\epsilon) \right)}{kj} &\leq -(\theta(k, \tau)-\tilde \epsilon). \label{equtheta}
\end{align}
For $m \geq kj$, $\beta(m,\tau,\epsilon) \leq \beta(kj,\tau,\epsilon)$. Hence,

\begin{align}
\limsup_{m \rightarrow \infty}  \frac{\log \left(\beta (m, \tau, \epsilon) \right)}{m} &\leq \limsup_{j \rightarrow \infty}  \frac{\log \left(\beta(kj,\tau,\epsilon) \right)}{kj}  \leq   - (\theta(k,\tau)- \tilde \epsilon). \notag
\end{align}
Note that the left hand side (L.H.S) of  the above equation does not depend on $k$. Taking supremum with respect to $k$ on both sides of the equation and noting that $\tilde \epsilon$ is arbitrary, proves $(i)$.
\end{IEEEproof}
\begin{remark}
Part $(ii)$ of Lemma \ref{lem:achandconverse} is known as the \textit{weak converse} for the HT problem in the literature, since it holds only when type 1 error probability tends to zero. Also, $(i)$ and $(ii)$ together imply that $\theta(\tau)$ is the optimal T2-EE as $\epsilon \rightarrow 0$, i.e., $\lim_{\epsilon \rightarrow 0} \kappa( \tau,\epsilon) = \theta(\tau)$.
\end{remark}
Part $(i)$ of Lemma \ref{lem:achandconverse} proves the achievability of the T2-EE $\theta(\tau)$  using Stein's Lemma. In Appendix \ref{appscheme}, we show an explicit proof of the achievability by computing the type 1 and type 2 errors for a block-memoryless stochastic encoding function at the observer and a joint typicality detector.

Note that for the  TACI problem,  the KL-divergence becomes mutual information, and we have
\begin{align*}
\theta(\tau) &= \sup_{\substack{f_1^{(k,n)}, \ldots,f_L^{(k,n)} \\ k, n \leq \tau k} } \frac{I(V^k;Y_{\Lcal}^n|Z^k)}{k}  \\
& \mbox{s.t. } (Z^k,V^k)-U_l^k-X_l^n=f_l^{(k,n)}(U_l^k)-Y_l^n, \forall~ l \in \Lcal 
\end{align*}

Although Lemma \ref{lem:achandconverse} implies that  $\theta(\tau)$ is an achievable T2-EE, it is in general not computable as it is defined in terms of a multi-letter  characterization. However,  as we will show below, for the TACI problem,  single-letter bounds for $\theta(\tau)$ can be obtained. By the memoryless property of the sequences $V^k$ and $Z^k$, we can write

\begin{align}
\theta(\tau) &= H(V|Z)-\inf_{\substack{f_1^{(k,n)}, \ldots,f_L^{(k,n)} \\ k, n \leq \tau k}} \frac{H(V^k| Y_{\Lcal}^n,Z^k)}{k}:   \label{compinf}  \\
&  (Z^k,V^k)-U_l^k-X_l^n=f_l^{(k,n)}(U_l^k)-Y_l^n ,~  \forall~ l \in \Lcal  \notag.
\end{align}

 In the next section, we introduce the $L-$helper JSCC problem  and show that the multi-letter characterization of this problem  coincides with obtaining the infimum in \eqref{compinf}. The computable characterization of the lower  and upper  bounds  for \eqref{compinf} then follows from the single-letter characterization of the $L-$helper JSCC problem.

\section{$L-$helper JSCC problem }

\begin{figure}[t]
\centering
\includegraphics[height=3.5cm, width=8cm]{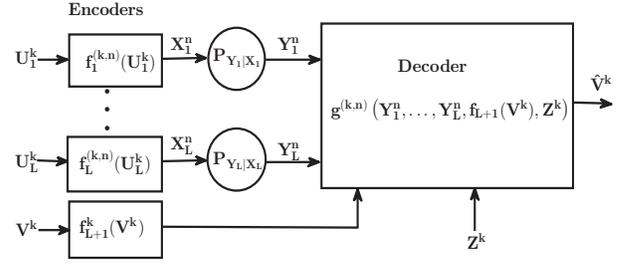}
\caption{$L-$helper JSCC problem.} \label{nhsc}
\end{figure}

Consider the model shown in Fig. \ref{nhsc} where there are $L+2$ correlated discrete memoryless sources $(U_{\Lcal},V,Z)$ i.i.d.  with joint distribution $P_{U_{\Lcal}VZ}$. For $~ 1 \leq l \leq L$, encoder $l$  observes the  sequence  $U_l^k$  and transmits   $X_l^n=f_l^{(k,n)}(U_l^k)$ over the  corresponding noisy channel, where  $f_l^{(k,n)}:\U_l^k \rightarrow \X_l^n$,  whereas encoder $L+1$ observes $V^k$, and outputs  $f_{L+1}^{k}(V^k)$, $f_{L+1}^{k}: \V^k \rightarrow \M=\{1,\ldots,2^{kR}\}$. The decoder has access to side-information $Z^k$, receives   $f_{L+1}^{k}(V^k)$ error-free, and also observes $Y_{\Lcal}^n$, the output of the DMCs $P_{Y_l|X_l}, ~ l \in \Lcal$.  The output of the decoder is given by the mapping $g^{(k,n)}: (\M,\Y_{\Lcal}^n,\Z^k) \rightarrow \hat V^k$. The decoder is interested in  reconstructing  $V^k$ losslessly. 
For a given bandwidth ratio $\tau$, a rate $R$ is said to be  achievable for the $L-$helper JSCC problem if  for every $\lambda \in (0,1]$,  there exist a  sequence of numbers $\delta_k \geq 0$ with $\lim_{k \rightarrow \infty} \delta_k =0$, encoders   $f_{L+1}^{k}(\cdot)$,  $f_l^{(k,n_k)}(\cdot),~ l \in \mathcal{L}$,  and decoder $g^{(k,n_k)}(\cdot,\cdot,\cdot)$, such that  $n_k \leq \tau k$ and   
\begin{align*}
 &\mathrm{Pr}\left(g^{(k,n_k)}\left(f^{k}_{L+1}(V^{k}),  Y_{\Lcal}^{n_k},Z^{k}\right) =  V^{k}\right)  \geq 1-\lambda,  \\
 &~~~~~~~~~~~ \text{ and } \frac{\log(|\M|)}{k} \leq R+\delta_k.
\end{align*}
The infimum of all achievable rates $R$ for the $L-$helper JSCC  problem with bandwidth ratio $\tau$ is denoted by $R(\tau)$.

Next, we  show that the problem of obtaining the infimum in \eqref{compinf} coincides with the multi-letter characterization of $R(\tau)$ for  the $L-$helper JSCC problem. 
Let

\begin{align}
R_{k} &\triangleq  \inf_{ \substack{f_1^{(k,n)},\ldots,f_L^{(k,n)} \\ n \leq \tau k}} \frac{H(V^k|Y_{\Lcal}^n,Z^k)}{k} \label{ratenhsc} \\
& \mbox{s.t. } (Z^k,V^k) -U_l^k- X_l^n= f_l^{(k,n)}(U_l^k)- Y_l^n,~ l \in \Lcal. \notag
\end{align}

\begin{theorem} \label{multiletternhsc}
For the $L-$helper JSCC problem, 
\begin{align*}
R(\tau)&= \inf_{k} R_{k}.
\end{align*}
\end{theorem}
\begin{IEEEproof}
The proof is given in Appendix \ref{thmproofnhsc}.
\end{IEEEproof}

 Having shown the equivalence between the multi-letter characterizations of $\theta(\tau)$ for the TACI problem over noisy channels and $R(\tau)$ for the $L-$helper JSCC problem, our next step is to obtain  computable single-letter lower and upper bounds on $R(\tau)$, which can then be used to obtain bounds on $\theta(\tau)$.
For this purpose, we use the   \textit{source-channel separation theorem} \cite[Th. 2.4]{Jin_Zhi}  for orthogonal multiple access channels. The theorem  states that all achievable average  distortion-cost tuples  in a multi-terminal JSCC (MT-JSCC) problem over an orthogonal multiple access
channel (MAC) can be obtained by the intersection of  the rate-distortion region and the MAC region. We need a slight generalization of this result when there is side information $Z$ at the decoder, which can be proved similar to  \cite{Jin_Zhi}.
Note that the $L-$helper JSCC problem is  a special case of the  MT-JSCC problem with $L+1$ correlated sources $P_{U_{\Lcal}V}$ and side information $Z$ available at the  decoder, where the objective is to reconstruct $V$ losslessly. 
 Although the above  theorem proves that separation holds, a single-letter expression is  not available in general for the multi-terminal rate distortion problem \cite{Elgamalkim}. However, single-letter inner and outer bounds have been given in  \cite{Elgamalkim}, which enable us to obtain single-letter  upper and lower bounds on $R(\tau)$  as follows.
\begin{theorem}
Let $C_l \triangleq \max_{P_{X_l}} I(X_l;Y_l), ~l \in \Lcal$ denote  the capacity of the channel $P_{Y_l|X_l}$, and $\tau$ the bandwidth ratio for  the $L-$helper JSCC problem. Define  
\begin{equation}
R^i(\tau) \triangleq  \inf_{W_{\Lcal}} \max_{S \subseteq \Lcal}  F_S,  \label{innerreg}
\end{equation}
where 
\begin{equation}
F_S=H(V|W_{S^c},Z)+I(U_S;W_S|W_{S^c},V,Z)- \tau \sum_{l \in S} C_l \notag
\end{equation}
for some auxiliary r.v.'s $W_l$, $l \in \Lcal$, such that  
\begin{align}
(Z,~V,~U_{l^c},~W_{l^c})-U_l-W_l, \label{btinnmrkv} 
\end{align}
 $|\mathcal{W}_l| \leq |\U_l|+4$, and for all subsets $S \subseteq \Lcal$,
\begin{align}
I(U_S;W_S|V,W_{S^c},Z) \leq  \tau \left( \sum_{l \in S} C_l \right). \label{innercond}
\end{align} 
 Similarly, let  $R^o(\tau)$ denote the right hand side (R.H.S) of \eqref{innerreg}, when the auxiliary r.v.'s $W_l,~l \in \Lcal$, satisfy  \eqref{innercond}, $|\mathcal{W}_l| \leq |\U_l|+4$ and  
 \begin{align}
  (V,U_{l^c},Z)-U_l-W_l.  \label{btoutmrkv}
\end{align}   
 Then,
\begin{align}
R^o(\tau) \leq ~&R(\tau) \leq R^i(\tau), \label{ratebnds} \\
H(V|Z)- R^i(\tau) \leq ~&\theta(\tau) \leq H(V|Z)- R^o(\tau). \label{thetabnds}
\end{align}
\end{theorem} 
\begin{IEEEproof}
From the  source-channel separation theorem, an upper  bound on $R(\tau)$  can be obtained   by the intersection of the Berger-Tung (BT) inner bound \cite[Th. 12.1]{Elgamalkim} with the capacity region $(C_1, \ldots, C_L, C_{L+1})$, where $C_{L+1}$ is the rate available over the noiseless link from the encoder of source $V$ to the decoder. Writing the BT inner bound  \footnote{ $R^i(\tau)$ can be improved by introducing a time sharing r.v. $T$ (independent of all the other r.v.'s) in the BT inner bound, but it is omitted  here for simplicity.} explicitly, we obtain that for all $S \subseteq \Lcal$ (including the null-set),
\begin{align*}
I(U_S;W_S|V,W_{S^c},Z) &\leq  \sum_{l \in S} \tau C_l,  \\
I(U_S;W_S|V,W_{S^c},Z) +H(V|W_{S^c},Z) &\leq \sum_{l \in S} \tau C_l +C_{L+1},
\end{align*}
where the auxiliary r.v.'s $W_{\Lcal}$ satisfy \eqref{btinnmrkv} and $|\mathcal{W}_l| \leq |\U_l|+4$.
Taking the  infimum of $C_{L+1}$ over all such $W_{\Lcal}$ and denoting it by $R^i(\tau)$,  we obtain  the second inequality in \eqref{ratebnds}. The other direction in \eqref{ratebnds} is obtained similarly by using the BT outer bound \cite[Th. 12.2]{Elgamalkim}.  Since $R(\tau)$ is equal to the infimum in \eqref{compinf}, substituting \eqref{ratebnds} in \eqref{compinf} proves \eqref{thetabnds}. 
\end{IEEEproof}

 The  BT inner bound is tight for the two terminal case, when one of the distortion requirements is zero (lossless) \cite[ Ch.12]{Elgamalkim}. Thus, we have the following result (for convenience, we drop the index 1 from the associated variables).
 
\begin{lemma} \label{corrsinguser}
For the TACI problem with $L=1$ and  bandwidth ratio $\tau$,
\begin{align}  
\theta(\tau)&= \sup_{W} I(V;W|Z) \label{thetsing}  \\
 \mbox{ such that }& I(U;W|Z) \leq \tau C, \label{capconstr}\\
 (Z,V)-U &-W, ~|\mathcal{W}| \leq |\U|+4 \label{markcond}
 \end{align}

\end{lemma}
\begin{IEEEproof}
Note that the Markov chain conditions in \eqref{btinnmrkv} and \eqref{btoutmrkv} are identical for $L=1$. Hence, 
\begin{align}
R^i(\tau)=R^o(\tau)=R(\tau). \label{ratetight}
\end{align}
 Using the BT inner bound in \cite[ Ch.12]{Elgamalkim}, we  obtain $R(\tau)$ as the infimum of $R'$ such that
\begin{align}
H(V|Z,W) &\leq R'\\
I(U;W|V,Z) &\leq \tau C \label{capconst2} \\
H(V|Z,W)+I(U;W|Z) &\leq \tau C+R'
\end{align}
for some auxiliary r.v. $W$ satisfying \eqref{markcond}. 
Hence,
\begin{align}
R(\tau)=& \inf_{W} \max \big( H(V|W,Z),~ H(V|W,Z) \notag \\
& \quad +I(U;W|Z)-\tau C \big)  \label{maxw}
\end{align}
such that  \eqref{markcond} and \eqref{capconst2} hold.
We next prove that \eqref{maxw} can be simplified as
\begin{align}
 R(\tau)= \inf_W H(V|Z,W) \label{finexprtau}
\end{align}
such that \eqref{capconstr} and \eqref{markcond} are  satisfied.
 This is done by  showing  that, for every r.v. $W$ for which  $I(U;W|Z)>\tau C$, there exists a r.v. $\bar W$ such that $I(U;\bar W|Z)=\tau C$, $H(V|\bar W,Z)\leq H(V| W,Z)+I(U;W|Z)-\tau C$ and \eqref{markcond} and \eqref{capconst2} are satisfied with $W$ replaced by $\bar W$. Setting
\begin{equation*}
\bar W =  
\begin{cases}
 W,  \quad \quad \quad \mbox{    with probability 1-p}, \\
  \mbox{constant}, \quad \mbox{with probability p}, 
 \end{cases}
\end{equation*}
 suffices, where we choose  $p$ such that $I(U;\bar W|Z)=\tau C$. The details can be found in \cite[Lemma 5]{Sree_ext}. Eqn. \eqref{thetsing} now follows from  \eqref{thetabnds}, \eqref{ratetight} and \eqref{finexprtau}.
  \end{IEEEproof}
\begin{remark}
We note here that the single-letter T2-EE characterization in Lemma  \ref{corrsinguser} exhibits a separation between the distributions of the data sources $U,V,Z$ and the channel distribution $P_{Y|X}$. Together with the fact that the  optimal $R(\tau)$ in the $L-$helper JSCC problem is achieved by separate source and channel coding, one might be inclined to assume that $\theta(\tau)$ for the TACI problem over noisy channels can also be achieved by a communication scheme that performs independent HT and channel coding, and the optimal T2-EE can be obtained by simply replacing the rate constraints in the TACI T2-EE expressions in \cite{Rahman-Wagner} with the corresponding channel capacity values. Although such a scheme is intuitively pleasing, the T2-EE analysis for such a scheme would involve a tradeoff between two competing error exponents, one being the T2-EE assuming that an error does not occur in channel decoding,  and the other  being   the  reliability function $E_r$ of the channel $P_{Y|X}$  \cite{Csiszar-Korner}. The details of the  analysis can be found in \cite{Sree_ext}.
\end{remark}

\section{Conclusions}
 We have  studied the T2-EE  for the distributed  HT problem over orthogonal noisy channels with side information available at the detector. For  the special case of TACI, single-letter upper  and lower bounds are  obtained for the T2-EE, which are shown to be tight when there is a single observer in the system. It is interesting to note that the reliability function  of the channel does not play a role in the T2-EE,  and a strict operational separation between HT and channel coding does not apply in general, even though the optimal T2-EE can be evaluated using the marginal distributions of the data  sources and the channels, rather than their joint distributions.  Obtaining  single-letter bounds for the general HT problem, and analyzing  the error exponents for the weighted sum of  the type 1 and type 2 errors  in the  Bayesian setting are some of the interesting problems for future research.

\begin{appendices} 

\section{ T2-EE using joint typicality detector } \label{appscheme}
Here, we provide the proof for the case $L=1$.
For given arbitrary integers $k$ and $n$ such that  $n \leq k \tau$, fix $f_1^{(k,n)}=P_{X_1^n|U_1^k}$. For any integer $j$ and sequence $u_1^{kj}$, the observer  transmits $X_1^{nj}= f_1^{(kj,nj)}(u_1^{kj})$ generated i.i.d. according to $\prod_{j'=1 }^j P_{X_1^n|U_1^k=u_1^{k}(j')}$. The detector declares $H_0:P_{U_1VZ}$ if  $\left(  Y_1^{nj},V^{kj},Z^{kj} \right) \in T_{[Y_1^nV^kZ^k]_{\delta_j}}^j $ (here $\delta_j \rightarrow 0$ as $j \rightarrow \infty$) and $H_1: Q_{U_1VZ}$ otherwise. 
 To simplify the exposition,  we denote $(Y_1^n,V^k,Z^k)$ and  $T_{[Y_1^nV^kZ^k]_{\delta_j}}^j$   by $W_{k,n}$ and $T_{[W_{k,n}]_{\delta_j}}^j$, respectively.  By the Markov lemma \cite{Elgamalkim}, the type 1 error probability  tends to zero as $j \rightarrow \infty$. The type 2 error probability is bounded by

\begin{align}
&\beta' \left(kj,nj, f_1^{(kj,nj)},\epsilon \right) \leq  Q_{Y_1^{nj}V^{kj}Z^{kj}}\left(T_{[Y_1^nV^kZ^k]_{\delta_j}}^j \right) \notag \\
&  \leq  \sum_{ \tilde P \in T_{[W_{k,n}]_{\delta_j}}^j} \sum_{w_{k,n}^j \in T_{\tilde P}} Q_{W_{k,n}^j}( w_{k,n}^j) \notag \\
& \firstequ \sum_{ \tilde P \in T_{[W_{k,n}]_{\delta_j}}^j} \sum_{w_{k,n}^j \in T_{ \tilde P}} 2^{-j \left(H(\tilde P)+D \left(  \tilde P||Q_{W_{k,n}} \right) \right)} \notag \\
& \secondequ \sum_{ \tilde P \in T_{[W_{k,n}]_{\delta_j}}^j }
 2^{-j D \left( \tilde P||Q_{ W_{k,n}} \right)}  \firstineq (j+1)^{|\W_{k,n}|} 2^{-j B(k,n) }  \notag
\end{align} 
where
\begin{align*}
B_{k,n}(j) \triangleq \min_{\tilde P \in T_{[W_{k,n}]_{\delta_j}}^j} D (  \tilde P||Q_{ W_{k,n}} ).
\end{align*} 
(a), (b) and (c) follow from Lemma's 2.3, 2.6  and 2.2 in \cite{Csiszar-Korner}, respectively.
Hence,
\begin{align*}
 \frac{\log\left(\beta' \left(kj,nj, f_1^{(kj,nj)},\epsilon \right)\right)}{kj}\leq -\frac{B_{k,n}(j)}{k}+ \delta_{k,n}'(j),
\end{align*}
where $\delta_{k,n}'(j) \triangleq \frac{|\W_{k,n}| \log(j+1)}{kj} $ and $|\W_{k,n}| \leq |\Y|^n|\V|^k|\Z|^k$. Note that for any $k$ and $n$, $\delta_{k,n}'(j) \rightarrow 0$ as $j \rightarrow \infty$.
Also, since  $\delta_j$ is chosen such that it tends to 0  as $j \rightarrow \infty$, $B_{k,n}(j)$ converges to $D(P_{W_{k,n}}||Q_{W_{k,n}})$ by the continuity of $ D (  \tilde P||Q_{ W_{k,n}} )$ in $\tilde P$ for fixed $Q_{ W_{k,n}}$. Since $k$, $n$ and  $f_1^{(k,n)}$  are  arbitrary, it follows from  \eqref{seqkappa} and \eqref{thetataudef} that  $\theta(\tau)$ is an  achievable T2-EE for any upper bound $\epsilon$ on the type 1 error probability.
 It is easy to see that this scheme  can be generalized to  $L>1$.

\section{Proof of Theorem \ref{multiletternhsc}} \label{thmproofnhsc}
For the achievability part, consider the following scheme.

\textbf{Encoding:}
Fix $k,n \in \mathbb{Z}^+$ and  $P_{X_l^n|U_l^k}$ at encoder $l$, $l\in \Lcal$. Let $j \in \mathbb{Z}^+$. On observing  $ u_l^{kj} $, encoder $l$ transmits  $X_l^{nj}=f_l^{(kj,nj)}(U_l^{kj})$  generated i.i.d. according to $\prod_{j'=1}^{j} P_{X_l^{n}|U_l^{k}=u_l^k(j')}$. Encoder $L+1$  performs uniform  random binning on  $V^k$, i.e,  $f_{L+1}^{kj}: \V^{kj} \rightarrow  \mathcal{M}=\{1, 2 , \cdots, 2^{kjR} \}$. By uniform random binning, we mean that $f_{L+1}^{kj}(V^{kj})=m$, where $m$ is selected uniformly at random from the set $\mathcal{M}$. 

\textbf{Decoding:} Let $M$ denote the received bin index, and $\delta >0$ be an arbitrary number.
If there exists a unique  sequence  $\hat V^{kj} $ such that $f_{L+1}^{kj}(\hat V^{kj})=M$ and $(\hat V^{kj},~Y_{\Lcal}^{nj},Z^{kj}) \in T_{[V^kY_{\Lcal}^nZ^k]_{\delta}}^j$, then the decoder outputs $g^{(kj,nj)}(M,Y_{\Lcal}^{nj},Z^{kj})= \hat V^{kj}$. 
Else, an error is declared. 

It can be shown that the probability of decoding error tends to $0$ as $j \rightarrow \infty$, if $R > H(V^k|Y_{\Lcal}^n,Z^k)+ \delta$. The details can be found in \cite[Appendix B]{Sree_ext}, along with the proof of the  converse.

\end{appendices}

\bibliographystyle{IEEEtran}
\bibliography{references}

\end{document}